\documentclass{sig-alternate-10pt}
\usepackage{graphicx}
\usepackage{amsmath}
\usepackage{bbm}
\usepackage{amssymb}
\usepackage{tikz}
\usepackage{subfigure}
\usetikzlibrary{arrows,automata}
\usepackage{empheq}
\usepackage{epstopdf}
\usepackage{helvet}
\usepackage{enumitem}
\usepackage{hyperref}
\hypersetup{backref}
\usepackage[all]{hypcap}

\begin{document}
\title{Towards Wire-speed Platform-agnostic Control of OpenFlow Switches}

\author{
Giuseppe Bianchi*, Marco Bonola*, Antonio Capone$^\dag$, Carmelo Cascone$^\dag$, Salvatore Pontarelli*\\
\affaddr{*CNIT/Universit\'a di Roma Tor Vergata  $^\dag$Politecnico di Milano}
}

\maketitle

\begin{abstract}
The possibility to offload, via a platform-agnostic specification, the execution of (some/part of the) control functions down to the switch and operate them at wire speed based on packet level events, would yield significant benefits in terms of control latency and reaction times, meanwhile retaining the SDN-type ability to program and instantiate a desired network operation from a central controller. While programmability inside the switches, at wire speed and using platform-independent abstractions, of ``any possible'' control function seems well beyond the OpenFlow capabilities, in this paper\footnote{
	This paper is an extended version of the work in progress \cite{ccr14}; extensions 
	involve i) hardware implementation, ii) several new use cases, iii) Mininet demonstration 
	including implementation of most use cases, iv) complete rewriting and restructuring of the 
	manuscript, and v) revised and updated related work.} 
we argue that a non trivial sub-class of stateful control functions, namely those that can be abstracted in terms of Mealy (Finite State) Machines, is {\em already} compatible with off-the-shelf OpenFlow version 1.1+ Hardware with marginal architectural modifications. With minimal additional hardware circuitry, the above sub-class can be extended to include support for bidirectional/cross-flow state handling. We demonstrate the viability of our proposed approach via two proof-of-concept implementations (hardware and software), and we show how some stateful control functionalities frequently exploited in network protocols are readily deployed using our application programming interface.
\end{abstract}



\section{Introduction}

Coined in 2009 \cite{Gre09}, the term {\em Software Defined Networking} (SDN) has gained significant momentum in the last years. SDN's promises to enable easier and faster network innovation, by making networks programmable and more agile, and by centralizing and simplifying their control. Even if some SDN's programmable networking ideas date back to the mid of the 90s \cite{Zeg14}, and do not nearly restrict to device-level programmability and to OpenFlow \cite{OF08}, it is fair to say that OpenFlow is the technology which brought SDN to the real world \cite{Gre09}. Quoting \cite{Zeg14}, ``{\em Before OpenFlow, the ideas underlying SDN faced a tension between the vision of fully programmable networks and pragmatism that would enable real-world deployment. OpenFlow struck a balance between these two goals by enabling more functions than earlier route controllers and building on existing switch hardware, through the increasing use of merchant-silicon chipsets in commodity switches}''.


{\bf Platform-independent forwarding abstraction}. The ``match/action'' programming abstraction of OpenFlow used for configuring the forwarding behavior of switching fabrics has the key feature of being {\em vendor-neutral}, arguably a crucial enabler for the SDN vision \cite{OF08}. This
%
abstraction brought about significant advantages in terms of both flexibility and simplicity; flexibility because the device programmer could broadly specify a flow via an header matching rule, associate forwarding/processing {\em actions} (natively implemented in the device) to the matching packets, and access statistics associated to the specified flow; simplicity because network administrators could handle heterogeneous multi-vendor network nodes, without the need to bother with a plethora of proprietary configuration interfaces.

{\bf Control and data plane separation}. The separation between control and data plane is highlighted as a distinguishing feature of SDN, and sometimes even postulated as the SDN definition itself. But should such a separation necessarily take the form of a {\em physical} separation, namely a ``smart'' controller (or network of controlling entities), which runs the control logic for ``dumb'' switching fabrics? This was the case with (the original) OpenFlow, as its ``match/action'' programmatic abstraction necessarily resorts on an external controller for (reactively or proactively) updating forwarding policies in the switches' flow tables. But the ability to {\em program} control functions directly inside the switch could lead to a different instantiation of the same {\em conceptual} control/data plane separation. A centralized controlling entity could {\em decide} how forwarding rules and flow states should change in front of network events, and {\em formalize} such a desired control logic via suitable platform-agnostic programming abstractions. The resulting control {\em programs} could then be not only centrally {\em executed} in the controller whenever they are triggered by network-wide events and require knowledge of network-wide states, but could also be {\em offloaded} down to the switching fabrics and {\em locally} executed, whenever they require only local states or events, leaving freedom to programmers of optimizing the mix of central and local/parallel processing capabilities. 

{\bf Programming control tasks inside the switch}. The ability to {\em execute} inside single links/switches, at wire-speed, control tasks triggered by packet arrivals or local measurements, would bring about significant advantages in terms of reaction time and reduced signalling load towards external controllers. Notable cases of where local state and forwarding behavior updates would be by far better handled locally, do include, but not nearly limit to: layer 2 MAC learning updates, request-response matches in bidirectional flows, multi-flow sessions or protocols (e.g. FTP’s control on port 21 and data on port 20), NAT states, packet fragmentation states, and so on. Indeed, reduced control latency and signalling overhead is a major motivation behind the emerging {\em hybrid switches} (which however resort on legacy control functions preimplemented in the switches), and behind some recent openflow {\em ad hoc} extensions (e.g. meters, fast failover). Rather than envisioning tailored control extensions, to the best of our knowledge, our proposal, OpenState, first challenges a more general question: how (and which) control functions can be programmed in the switch, while guaranteeing i) platform independent programming abstractions, ii) wire speed operation, and iii) reuse of existing OpenFlow commodity hardware and currently available OpenFlow action set? Note that we {\em do not} claim, of course, the ability to program any possible control function, as this would require supplementary actions and further extensions; rather, our target is to understand which subset of control tasks can be formally described using platform-independent abstractions, and how our proposed abstractions can be supported into an OpenFlow switch with minimal modifications.

{\bf Contribution}. The approach proposed in this paper, which we descriptively call {\em OpenState}, focuses on the introduction of {\em programmable states and state transitions in OpenFlow}. We model control logic as {\em the ability to exploit packet-level events to trigger wire-speed changes in the forwarding plane rules and the relevant flow states}. Our main results summarize as follows.
\vspace{-5pt}
\begin{itemize}\setlength{\itemsep}{-3pt} 
\item Any control logic (relying on the stateful invocation of already available OpenFlow actions) which can be described in the abstract (platform-agnostic) form of a Mealy Finite State machine, can be readily supported on OFv1.3+ switches with marginal amendments, and can be executed at wire speed. 
\item With minimal hardware extensions, OpenState can support ``cross-flow'' state handling (arrival of a packet of a given flow triggering a state transition for a different flow - as e.g., needed in MAC learning, frames forwarded based on destination MAC, but states being updated based on source MAC).
\item We prove OpenState's viability by i) showing that it requires minimal software modifications in an existing (softswitch) OpenFlow version 1.3 implementation \cite{ofsoftswitch13}, and by ii) proving via a proof of concept hardware implementation (minimal in scale and features, but sufficient to identify implementation issues and limitations), that it can operate at a (wire) speed comparable to OpenFlow. 
\item By means of simple use case examples, we show how small control tasks, frequently recurring as building blocks for more comprehensive protocols, can be readily supported over OpenState. Most of the use cases (as well as others not presented here for space reasons) are implemented on Mininet and are publicly avaliable for testing at \cite{openstate-homepage}, along with the OpenState softswitch implementation.
\end{itemize}



\section{Approach sketch} 
\label{s:sketch}

OpenState's goal is to entail platform-agnostic programming of per-flow (broadly defined as an OpenFlow match) control logic directly within network nodes, thus rescinding the today's necessary reliance on external (slow-path) controllers. For the purposes of this paper, we operatively define control logic as the {\em ability to install or update, at wire speed (on a per packet basis) a flow table entry associated to a flow and a relevant flow state, on the basis of packet level events, such as specific header matches, and the current flow state information}. Since, in this work, we look for a programmatic abstraction, similarly to \cite{OF08}, we will remain agnostic to the specific set of actions that can be associated to a flow table match. Of course, support for some practical control functions require extensions in the set of actions supported by the switch; such extension is out of the scope of the present paper, and we defer the reader to the conclusive discussion. 

Although obviously restrictive with respect to all (!) the possible flow-level control tasks that may be envisioned, the above operative definition entails a level of flexibility in the switch which goes well beyond what is currently supported in today’s OpenFlow switches. Indeed, OpenFlow-type forwarding rule configurations are static inside the switch, consistently apply to all packets of a flow irrespective of what ``has happened'' with the previous packets, and are changed only upon intervention of an external controller entity. Conversely, OpenState aims at providing the possibility to configure, and dynamically enforce at wire speed and within the switch, i) programmer-specific (custom) states which determine which filtering/forwarding rules should be applied, and ii) a formal, programmer-specific, behavioural description of how flow states shall evolve, i.e., which packet arrivals or events should trigger state transitions.

{\bf A pragmatic approach}. The history of programmable networks \cite{Zeg14} reports many cases where technically compelling solutions (such as the IETF ForCES protocol, or some active networking paradigms) have ultimately failed to significantly impact the real world. Indeed, the OpenFlow success story suggests that pragmatism and viability are perhaps even more important than technical breakthroughs in flexibility and programmability, and that a careful balance must be done between a fully programmable vision and approaches which compromise in generality, but enable real world deployment. This balance was arguably found in the practical, but at the same sufficiently flexible, ``match/action'' primitive at the basis of the OpenFlow operation. The subsequent evolution in the OpenFlow specification (version 1.1 and beyond) has promoted many major extensions (pipelined flow tables, action bundles, synchronized tables, more flexible header matching, new actions, dedicated flow control structures such as meters, etc). Nevertheless, no substantial changes in the original programmatic abstraction have emerged in the mean time, for the likely reason that match/action rules are directly mapped over TCAM entries, unlike alternative abandoned proposals (such as Google's OpenFlow 2.0 \cite{Mey13}) which might require a non trivial departure from the consolidated OpenFlow switch architecture and its relevant commodity hardware components. 

Getting to OpenState' specificities, this brings us to the need to devise a programmatic approach which, despite challenging a very different and broadened goal (programmability of stateful control tasks) with respect to OpenFlow (stateless forwarding plane configurations), should be essentially deployable with very limited modifications to existing switch implementations, be they HW or SW ones. Especially in the case of HW switches, our approach should prove to be compatible with the existing commodity hardware (e.g., TCAMs) used in commercial devices, so that any required modification should essentially reduce to easy-to-perform firmware upgrades. Finally, and in all (SW or HW) cases, our promoted approach by no means should require vendors to open their implementation’s internals. 


\subsection{From match/action to Mealy machines}
\label{s:mealy}

OpenState builds on a (perhaps surprising) finding: the {\em very same} OpenFlow ``match/action'' primitive can be reused for a different goal and with a broadened semantic. In OpenFlow, a forwarding action (action set) is associated to a flow match. Taking advantage of the possibility, available since OpenFlow version 1.1, to extend the match to further metadata and associate more than one action (and/or instruction) to the outcome of said match, OpenState proposes to i) perform matches on packet header fields {\em plus a flow state label} (to be retrieved as discussed in $\mathsection$ \ref{s:state}), and ii) associate to such match both a forwarding action (or set of actions) {\em and a state transition}.
%
%
. Note that a match not triggering any state transition (arguably the most common case) is readily accounted in OpenState under the special case of {\em self-transitions}, i.e. a transition from a state to itself. 

The proposed approach can be formally modeled, in abstract form, by means of a simplified type of eXtended Finite State Machine (XFSM \cite{Che93}), known as {\em Mealy Machine}. We recall that a Mealy Machine is an abstract model comprising a 4-tuple $<S, I, O, T>$, plus an initial starting (default) state S0, where:
\vspace{-5pt}
\begin{itemize}\setlength{\itemsep}{-3pt} 
\item $S$ is a finite set of states; 
\item $I$ is a finite set of input symbols (events);  
\item $O$ is a finite set of output symbols (actions); and 
\item $T : S \times I \rightarrow S \times O$ is a transition function which maps $<$state, event$>$ pairs into  $<$next\_state, action$>$ pairs. 
\end{itemize}
Similarly to the OpenFlow API, the abstraction is made concrete (while retaining platform independency) by restricting the set $O$ of actions to those available in current OpenFlow devices, and by restricting the set $I$ of events to OpenFlow matches on header fields and metadata easily implementable in hardware platforms. The finite set of states S (concretely, state labels, i.e., bit strings), and the relevant state transitions, in essence the ``behavior'' of a stateful application, are left to the programmer's freedom. As discussed in $\mathsection$ \ref{s:archi}, a transition function $T$ is readily accomodated into a {\em single} TCAM entry, hence it uses the same OpenFlow hardware employed for ordinary match/action pairs. 

\begin{figure}[t]
\centering
\includegraphics[width=\columnwidth]{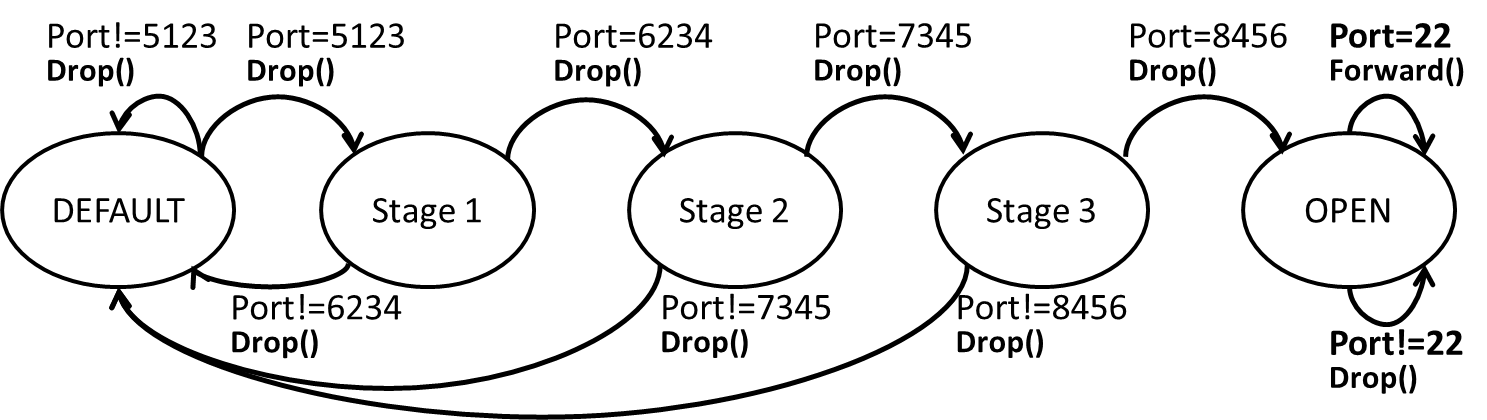}
\caption{port knocking example: Mealy (Finite State) Machine}
\label{fig:pk:mm}
\vspace{-0.4cm}
\end{figure}

{\bf An illustrative example: Port Knocking}. To best convey our concept, let's reuse (from \cite{ccr14}) a perhaps niche, but indeed very descriptive example: port knocking, a well-known method for opening a port on a firewall. An host IP that wants to establish a connection (say an SSH session, i.e., port 22) delivers a sequence of packets addressed to an ordered list of pre-specified closed ports, say ports 5123, 6234, 7345 and 8456. Once the exact sequence of packets is received, the firewall opens port 22 for the considered host. Before this stage, all packets (including the knocking ones) are dropped. This example is readily modelled via the Mealy Machine illustrated in Fig. \ref{fig:pk:mm}. Starting from a DEFAULT state, each correctly knocked port will cause a transition to a series of three intermediate states, until a final OPEN state is reached. Any knock on a different port will reset the state to DEFAULT. When in the OPEN state, only packets addressed to port 22 will be forwarded; all remaining packets will be dropped, but without resetting the state. Note that a controller-based implementation of Port Knocking would require the switch to deliver {\em each and every packet received on a currently blocked port} to the controller itself!

\subsection{Processing flow states}
\label{s:state}

Although the Mealy Machine provides an abstraction to {\em program} a desired state evolution along with associated forwarding actions, it does not say anything about {\em which flow entity} is assigned such state (for instance, the specific host IP address in the above port knocking example).  The state associated to a flow entity shall be stored in a separate flow table, and an arriving packet is thus handled via three steps: 
\begin{enumerate}\setlength{\itemsep}{-2pt} 
\item {\bf State lookup} - a query to a table, using as key a ``flow identity'', which returns a (unique) associated state label (DEFAULT if flow not found);
\item {\bf Extended finite state machine execution} - a Mealy Machine state transition step: given, as input, the retrieved state label and the packet header fields which specify an event match, it returns the label of the next state and triggers the forwarding action(s) associated to the state transition;
\item {\bf State update}: the retrieved (next) state label is {\em rewritten} in the state table for the relevant flow entity (or a new entry is added to the state table). 
\end{enumerate}

\begin{figure}[t]
\centering
\includegraphics[width=0.9\columnwidth]{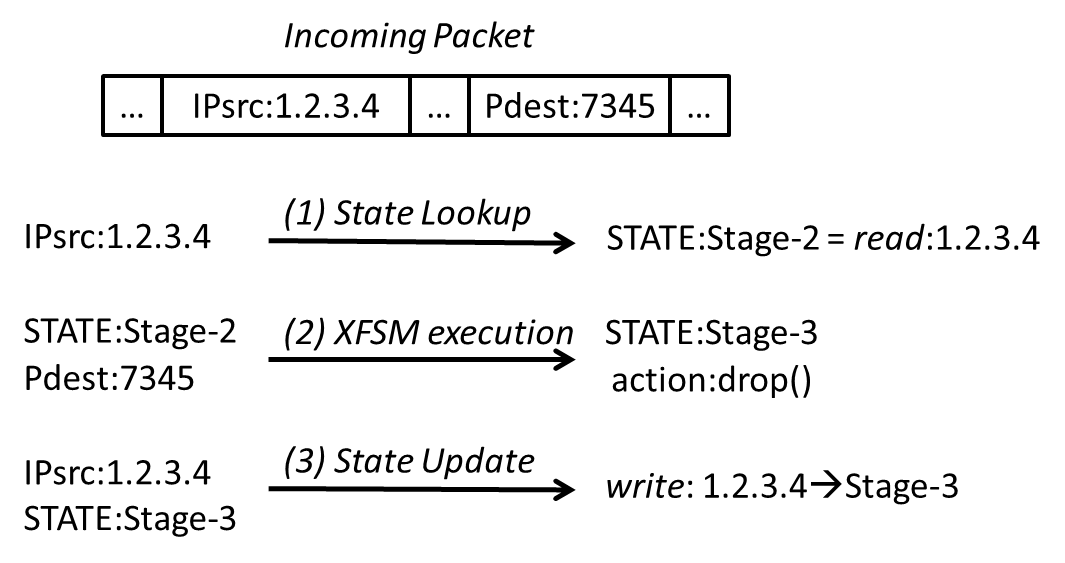}
\caption{port knocking example: Processing of a packet}
\label{fig:pk:ex}
\vspace{-0.4cm}
\end{figure}

It is worth to remark that the above operation clearly distinguishes the packet header matches exploited during step (2), which define the {\em events} triggering state transitions and which are associated to the forwarding action (or action set), from those used in steps (1) and (3), which define the {\em flow identities}, meant as entities which are attributed a state. Fig. \ref{fig:pk:ex} exemplifies such three steps for the port knocking state machine, assuming an incoming packet from the flow identity 1.2.3.4 (the host IP) which causes a port 7345 match event.

For a perhaps extreme, but hopefully descriptive, analogy, the above system acts as a sort of ``flow state processor''. The first step is in charge to identify the context (the specific flow identity found in the packet) and fetch the associated state, the second is the processing step, and the third step saves the resulting state and leaves the system ready to process the next packet. Note that the ``micro-program'', which determines how a flow-state should be processed, is {\em sofware-coded} in the abstract form of a finite size list of Mealy Machine transitions, it {\em does not change} during run time operation, and it is {\em shared by a potentally huge number of flows} (all possible source IP addresses arriving to the switch, in the port knocking example). 

This consideration has an important practical consequence. A Mealy Machine implementation requires wildcard matches and hence must be stored in a (costly) TCAM as discussed in the next $\mathsection$ \ref{s:archi}. Conversely, state lookups and updates are performed over flow identities and do not require wildcard matches (or require very few wildcard entries to manage exceptions, see $\mathsection$ \ref{s:hw:proto}); therefore flow states can be saved in an ordinary RAM-based hash table.  Since the number of states and events associated to states (i.e. the number of mealy machine transitions comprising the ``control program'') is, in general, much smaller than the number of flows controlled by such program, our proposed approach uses TCAM resources very sparingly, and with no need for (wire-speed) dynamic updates - dynamic changes occur only in the table storing the flow states.
 
\subsection{Cross-flow state management}
\label{s:xstate}

The above discussed decoupling between flow identities and matching events brings about a further, more subtle, advantage: it permits to support a functional extension which we call {\em ``cross-flow'' state handling}. There are many practically useful stateful control tasks, in which states for a given flow are updated by events occurring on {\em different} flows. A prominent example is MAC learning: packets are forwarded using the {\em destination} MAC address, but the forwarding database is updated using the {\em source} MAC address. Similarly, the handling of bidirectional flows may encounter the same needs; for instance, the detection of a returning TCP SYNACK packet could trigger a state transition on the opposite direction. In protocols such as FTP, a control exchange on port 21 could be used to set a state on the data transfer session on port 20. And in reverse path forwarding, a node forwards a packet based on destination, but must store a state for the return path.

To support such tasks, it suffices to conceptually separate the {\em identity} of the flow to which a state is associated, from the actual {\em position} in the header field from which such an identity is retrieved. This is accomplished by providing the programmer with the ability to differentiate a so-called {\em lookup-scope} from the {\em update-scope}, devised to identify the packet header fields used to extract from the packet the flow key used to access the state table during the state lookup and the state update stages, respectively. In order to support such an additional feature, the OpenFlow architecture must be extended with additional (simple) hardware implementing a {\em key extractor} circuitry ($\mathsection$ \ref{s:hw:proto}) which retrieves different packet header bytes during the different lookup and update stages. As an example, MAC learning is trivially programmed as described in $\mathsection$ \ref{s:mac-learning-uc}. 



\section{OpenState}
\label{s:archi}

\subsection{Application Programming Interface}
\label{s:api}

\begin{figure}[t]
\centering
\includegraphics[width=0.73\columnwidth]{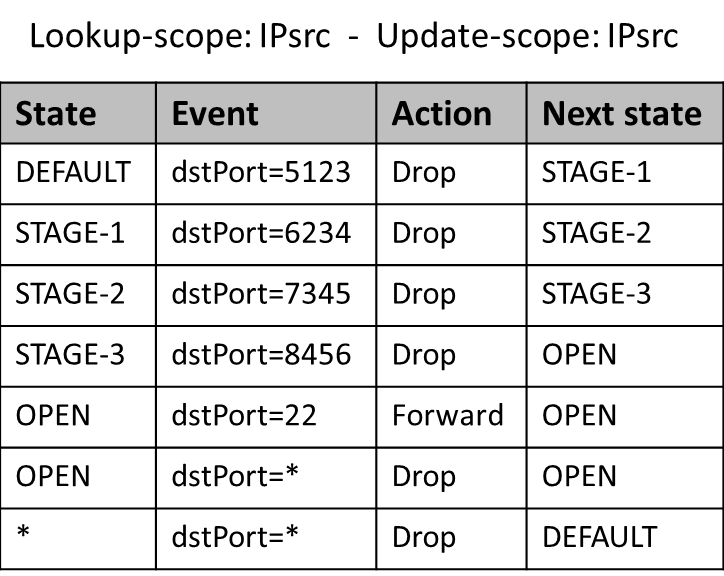}
\caption{port knocking example: control program - matches assumed to be performed in order (earliest match returned)}
\label{fig:pk:prg}
\vspace{-0.2cm}
\end{figure}

The OpenFlow data plane abstraction is based on a single table of match/action rules for version 1.0, and multiple tables from version 1.1 on. Unless {\em explicitly} changed by the remote controller through flow-mod messages, rules are {\em static}, i.e., all packets in a flow experience the same forwarding behavior. With OpenState, we introduce the notion of {\em stateful block}, as an extension of a {\em single} flow table. Stateful blocks can be pipelined with other stateful blocks as well as ordinary OpenFlow tables. A stateful block is an atomic block comprising two distinct, but interrelated, tables:
\vspace{-5pt}
\begin{itemize}\setlength{\itemsep}{-2pt}
\item a {\bf State Table}, which stores the state labels associated to flow identities (no state stored meaning DEFAULT state), and
\item an {\bf XFSM table}, which performs a (wildcard) match on a state label and the packet header fields, and returns an associated forwarding action (action set) and a next state label.
\end{itemize}
The programmer can specify the operation of a stateful block as follows:
\vspace{-5pt}
\begin{itemize}\setlength{\itemsep}{-2pt}
\item provide the list of entries to be loaded in the XFSM table. Each entry in the XFSM table comprises four columns: i) a \textbf{state} provided as a user-defined label, ii) an \textbf{event} expressed as an OpenFlow match, iii) a list of OpenFlow \textbf{actions}, and iv) a \textbf{next-state} label; each row is a designed state transition. At least one entry in the XFSM table must use, in the first column, the DEFAULT state label;
\item provide a ``lookup-scope'', namely the header field(s) of the packet which shall be used to access the state table during a lookup (read):
\item provide a possibly different ``update-scope'',  namely the header field(s) of the packet which shall be used to access the state table\footnote{
	As outlined in the usecase presented in $\mathsection$ \ref{s:forwarding-uc} a straightforward extension could consist in what follows. Rather than using a 
	{\em common} (unique) update scope for all the XFSM entries, an extended implementation 
	could permit the programmer to specify {\em different} update-scopes to be used as a parameter when calling for a state transition.}
\end{itemize}
As an example, figure \ref{fig:pk:prg} shows the ``program'' corresponding to the port knocking example discussed in section \ref{s:mealy}: the lookup-scope is of course equal to the update-scope and it is set to the source IP address. 

\subsection{Hardware viability}

\begin{figure*}[t]
\centering
   \includegraphics[width=0.9\textwidth]{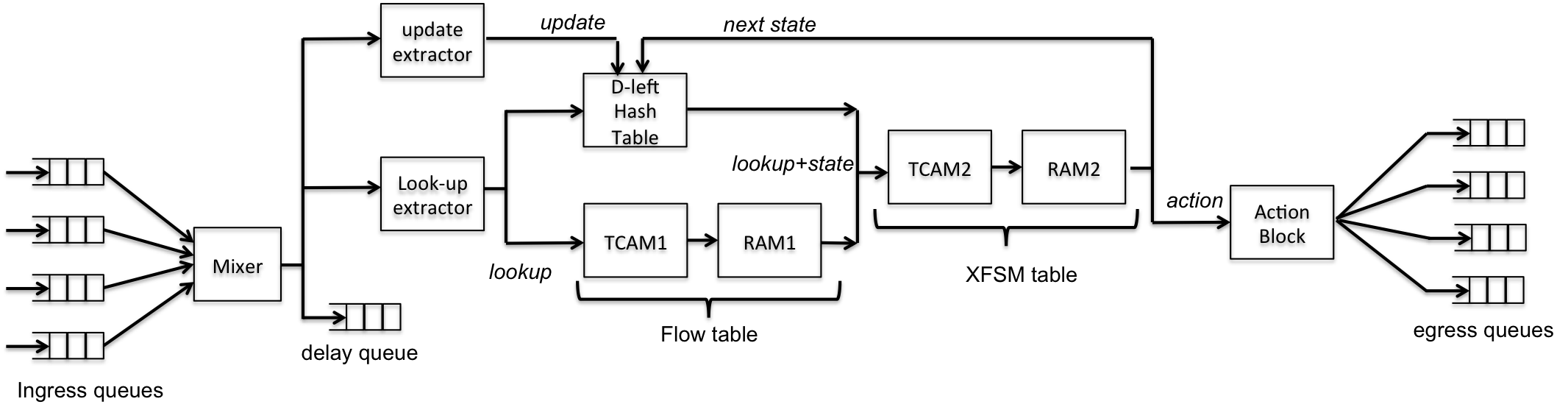}
\vspace{-0.3cm}
\caption{Scheme of the OpenState hardware prototype}
\label{F:archi}
\vspace{-0.3cm}
\end{figure*}

To gain understanding on the feasibility and wire-speed operation of OpenState, we have implemented a proof-of-concept hardware prototype using an experimental FPGA platform. The designed hardware prototype is conservative in terms of TCAM entries and clock frequency\footnote{
	Effective and scalable FPGA implementation of Ternary Content Addressable Memories is a widely open 
	research issue \cite{TCAM1,TCAM2,TCAM4}, especially since the priority resolution hardware limits the 
	maximum operating frequency when the number of TCAM entries increase. Indeed, the 
	achievable performance with FPGA TCAM are still far, in terms of size and clock 
	frequency, from those attainable by a full custom ASIC TCAM design \cite{TCAM3}. Further pushing 
	our FPGA implementation is thus not only well out of the scope of this work, but it is also of limited 
	practical relevance, as carrier-grade implementations in the order of Terabit/s throughput \cite{Bos13} 
	in any case would require a custom ASIC implementation.} 
but it includes all the key OpenState components and features, including support for cross-flow state management. Considering that the TCAM size does not affect throughput given its O(1) access time, we believe that even if at reduced scale, our implementation permits us to comparatively argue about complexity and performance with respect to an equivalent OpenFlow implementation. 

\subsubsection{FPGA prototype}
\label{s:hw:proto}

{\bf Development platform}. The OpenState hardware prototype has been designed using as target development board the INVEA COMBO-LXT \cite{invea}, an express PCI x8 mother card equipped with the XILINX Virtex5 XC5VLX155T \cite{V5}, two QDR RAM memories with a total capacity of 9MB and a throughput of 17166 Mbps for read and for write operations, and up to 4 GB of DDR2 memory. The 2 QDR II SRAM chips provide high bandwidth dual port memory for routing tables, flow memory, low latency data buffers. The board is equipped with a daughter board providing two 10 GbE interfaces, and is hosted in a PC workstation. The operating system sees the board as two Ethernet network cards connected using the express PCI bus. This allows to configure the development board as a 4 port switch in which two ``virtual'' ports are connected to the host workstation, while the other two ports can be connected to the external environment. All the four ports are connected to the FPGA that implements the OpenState architecture. 

{\bf Node prototype}. The general scheme of the OpenState prototype (single stateful block) is depicted in figure \ref{F:archi}. The FPGA is clocked at 156.25 MHz, with a 64 bits data path from the Ethernet ports, corresponding to a 10 gbps throughput per port. Four {\em ingress queues} collect the packets coming from the ingress ports. A 4-input 1-output {\em mixer block} aggregates the packets using a round robin policy. The output of the mixer is a 320 bits data bus able to provide an overall throughput of 50 Gbps. A {\em delay queue} stores the packet during the time need by the OpenState tables to operate.

{\bf State table}. 32 bit state labels are stored in a {\em d-left hash table}, with d=4. The table is sized for 4K entries, and is accessed with 128 bit keys provided, alternatively, by a {\em look-up extractor} and an {\em update extractor} during state read and (re)write, respectively. Two operations have been implemented for the lookup/update-scope: the first operation selects the beginning of the header, while the second operation is a 128 bits configuration mask register that individually mask the header bits. Two configuration registers have been used to configure the shifter and the mask. The barrel shifter takes as input the 320 bits data bus and provide as output 128 contiguous bits starting from the i$th$ bit defined by the first configuration register. The mask operation is simply the bitwise 'and' between the output of the barrel shifter and the second configuration mask register. The state table is further equipped with a TCAM with 32 entries of 128 bits (TCAM1), and an associated Block RAM of 32 entries of 32 bits (RAM1) that reads the output of the TCAM and provides the state associated to the specific TCAM row. The TCAM 
is needed to handle special (wildcard) cases, such as static state assignment to a pool of flows (e.g. ACLs), flow categories which are out of the scope of the machine and must be processed in a different way (if necessary by another stage, either stateless or stateful), etc. 

{\bf XFSM table}. The XFSM table is implemented by a TCAM with 128 entries of 160 bits (TCAM2), associated to a Block RAM of 128 entries of 64 bits (RAM2), storing the next state used to update the flow table, and the specific action to perform on the packet. This TCAM  takes as input the aggregated (unmasked) lookup-scope and the retrieved flow state, providing as output row associated to the matching rule with higher priority. As previously mentioned, the limited number of entries of the TCAMs is due to the inefficient mapping of these structures on an FPGA. This number, however, is similar to that of other FPGA based TCAM implementations, such as \cite{OpenFlowFPGA}. 

{\bf Packet output}. A final {\em Action Block} applies the retrieved action to the packet coming from the delay queue. Being our prototype a proof-of-concept, as of now only a basic subset of OpenFlow actions have been implemented: drop, select (enable one or more of the output ports to forward the packet), and tag (set a packet field to a specific value). This block then provides as output the four 64 bits data-bus for the four 10 Gbits/sec egress ports.

{\bf Synthesis results}. The whole system has been synthesized using the standard Xilinx design flow: the resource occupation for the implemented system, in terms of used logic resources, are presented in the table below. 

\noindent
\begin{tabular}{|c|c|c|}
  \hline
type of resource & \# of used resources & [\%] \\
\hline
Number of Slice LUTs & 10,691 out of  24,320  &  43\%   \\
 \hline
Block RAMs  &	 53 out of     212   & 25 \% \\
  \hline
\end{tabular}

\subsubsection{Discussion, Limitations, Extensions}

The FPGA prototype confirms the feasibility of the OpenState implementation. The additional hardware needed to support cross-flow state management (namely, the extractor modules) uses a negligible amount of logic resources and does not exhibit any implementation criticality. Similarly, the limited number of actions and TCAM entries  implemented in the prototype are just due to the proof-of-concept nature of our prototype (and lack of an OpenFlow hardware from which OpenState would directly inherit these parts). 

{\bf Limitations}. If compared with an OpenFlow implementation, OpenState exhibits only one (minor) shortcoming. The system latency, i.e. the time interval from the first table lookup to the last state update is 5 clock cycles. The FPGA prototype is able to sustain the full throughput of 40 Gbits/sec provided by the 4 switch ports. If we suppose a minimum packet size of 40 bytes (320 bits), the system is able to process 1 packet for each clock cycle, and thus up to 5 packets could be pipelined. However, the feedback loop (not present in the forward-only OpenFlow pipelines \cite{OF1.4}) raises a concern: the state update performed for a packet at the fifth clock cycle would be missed by pipelined packets. This could be an issue for packets {\em belonging to a same flow} arriving back-to-back (consecutive clock cycles); in practice, as long as the system is configured to work by aggregating $N \geq 5$ different links, the mixer's round robin policy will separate two packets coming from the same link of $N$ clock cycles, thus solving the problem. Note that the 5 clock cycles latency is fixed by the hardware blocks used in the FPGA (the TCAM and the Block RAMs) and basically does not change scaling up the number of ingress ports or moving to an ASIC.

{\bf Performance achievable with an ASIC implementation}.
As previously stated, while an FPGA prototype permits to assess feasibility, a full performance/scale architecture requires ASIC technology. Following the same technology assumptions of \cite{Bos13}, an OpenState ASIC design 
would be able to work at 1GHz operating frequency. This corresponds to an aggregate throughput of 960M packets/s, that is the maximum achievable by a 64 ports 10 Gb/s switch chip. However, the most important scaling provided by the ASIC implementation is given by the number of entries that can be stored in the OpenState tables. The size of the SRAM that can be instantiated on a last generation chip is up to 32 MB, corresponding to 2 millions of entries in the d-left hash for the Flow table. The size of a TCAM can be up to 40 Mb, corresponding to 256K XFSM table entries.



\subsection{Software viability}

To support our thesis that OpenState is readily implemented as a simple OpenFlow extension, we have developed a SW implementation by extending the OpenFlow 1.3 software switch \cite{ofsoftswitch13}, chosen because of its simplicity and adherence to the OpenFlow v1.3 specification, with our proposed stateful operation support. Our implementation, available at \cite{openstate-homepage}, comprises: (i) a userspace Openstate datapath implementation; (ii) a Openstate controller implementation; (iii) a set of runnable use case application examples (in part described in Section \ref{s:usecases}); (iv) a mininet distribution supporting our datapath and controller implementations. 

{\bf Supplementary Features}. The software implementation includes some supplementary details with respect to the OpenState operation described so far. Two possible state labels are returned in the case of a state table lookup miss: the ordinary \texttt{DEFAULT} state if the flow identity key is not found in the table, or a special \texttt{NULL} state value if the header fields specified by the lookup-scope are not found in the packet (e.g. extracting the IP source address when the Ethernet type is not IP). State transitions are supported by a supplementary command developed as an OpenFlow instruction, specifically a new \emph{SET\_STATE} instruction that will immediately trigger an update of the previous state table.The usage of an instruction guarantees that the state update is performed at the end of the stateful block, even when action bundles are configured, and permits to pipeline our stateful stage with supplementary stages, including other stateful ones. Finally, configurable timeouts have been added to manage soft states and add support for 'timer-based' events (a state transition is triggered by a timer expiration). 

{\bf OpenState datapath details}. To support advertisement and configuration of the proposed state management, a new capability bit \texttt{OFPC\_TABLE\_STATEFULL} and a new table configuration \texttt{OFPCT\_TABLE\_STATEFULL} bit have been defined. If \texttt{OFPCT\_TABLE\_STATEFULL = 1}, 
right after the packet headers are parsed, the flow state is retrieved and written as packet metadata, otherwise the packet directly jumps to the flow table. The basic flow table data structure has been extended with a support data
structure implementing the state table (a hash map indexed by the flow key) and the lookup and setup key extractor (two ordered lists of flow match TLV field indexes). The state table entries consists of the following fields:

\vspace{-5pt}
{\begin{footnotesize}
\begin{verbatim}
struct ofp_state_entry {
    uint32_t key_len;                                                      
    uint32_t state;
    uint8_t key[OFPSC_MAX_KEY_LEN];                                        
    uint32_t timeout;
    uint32_t to_state;                                                     
};                          
\end{verbatim}
\end{footnotesize}
}

\vspace{-5pt}
\noindent
A new OpenFlow instruction \texttt{OFPIT\_SET\_STATE} has been added to allow the
OpenFlow extended datapath to update the next state for a given flow.

\vspace{-5pt}
\begin{footnotesize}
\begin{verbatim}
struct ofp_instruction_set_state {
    uint16_t type; /* OFPIT_SET_STATE */
    uint16_t len;  /* Length is 16. */
    uint32_t state; /* Next state value */
    uint32_t timeout; /* State soft timeout [us] */
    uint32_t to_state; /* Roll back state value*/
};
\end{verbatim}
\end{footnotesize}

\vspace{-5pt}
\noindent
Moreover, a modify messages called \texttt{OFP\_STATE\_MOD} have been defined along
with the relevant message structure to allow a OpenFlow controller to
respectively configure the flow state entries and the key extractors:

\vspace{-3pt}
\noindent
\begin{footnotesize}
\begin{verbatim}
enum ofp_state_mod_command {
    OFPSC_SET_L_EXTRACTOR = 0,
    OFPSC_SET_U_EXTRACTOR,
    OFPSC_ADD_FLOW_STATE,
    OFPSC_DEL_FLOW_STATE
};
\end{verbatim}
\vspace{-7pt}
\begin{verbatim}
struct ofp_state_mod {
    struct ofp_header header;
    uint64_t cookie;
    uint64_t cookie_mask;
    uint8_t table_id;
    uint8_t command; //ofp_state_mod_command
    uint8_t payload[];
};
\end{verbatim}
\end{footnotesize}

\vspace{-5pt}
\noindent
No further (meaningful) modications were needed, the rest of the implementation relies on the standard OpenFlow data and message structures. 

\begin{figure}
	\centering
	\includegraphics[width=0.5\textwidth]{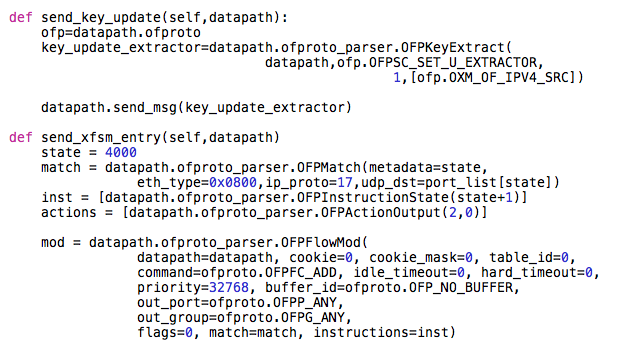}
	\caption{Excerpt of a Openstate application written in python}
	\label{f:ryu}
	\vspace{-0.3cm}
\end{figure}

{\bf Openstate controller}.
To support Openstate features at control plane, the well known Ryu SDN software framework \cite{RYU} has been extended in order to integrate the \texttt{OFPT\_STATE\_MOD} message and the \texttt{OFPIT\_SET\_STATE} instruction. The message structure for both messages reflects the \texttt{ofp\_state\_mod} and \texttt{ofp\_instruction\_set\_state} structures defined above. Figure \ref{f:ryu} shows an excerpt of an OpenState application code. 

\section{Example OpenState Applications}
\label{s:usecases}
In this section, with no pretense of completeness, simple OpenState control programs, are described. These applications are not nearly claimed to be innovative, and indeed they are frequently found as building blocks in literature proposals; rather, our goal is to show the versatility of OpenState in permitting the deployment of an heterogeneous subset of network functionalities which cannot be, as of now, implemented in OpenFlow swicthes without either extensions or the explicit intervention of external controllers.

\subsection{Learning}
\label{s:mac-learning-uc}

As duly anticipated, MAC learning, a well known OpenFlow limitation frequently invoked to motivate OpenFlow extensions \cite{Cra13}, becomes trivial by exploiting OpenState's cross-state management facility, see below. Moreover, OpenState does {\em not} implement learning as an internal extended action or control task, specific to MAC addresses (unlike the case of hybrid switches), but leaves complete freedom to the programmer to decide what should be learned and by which flow identifiers - see the discussion at the end of the section about the possibility to ``learn'' MPLS labels. 

{\bf MAC learning}. To configure OpenState as a MAC learning switch, it suffices to send the following commands from the controller:

\vspace{-6pt}
\begin{enumerate}\setlength{\itemsep}{-1pt} 

\item Set the {\em lookup-scope} to the Ethernet destination address, via an \texttt{OFPT\_STATE\_MOD} with command field set to \texttt{OFPSC\_SET\_L\_EXTRACTOR}, field count set to 1, and field list set to \texttt{[OFPXMT\_OFB\_ETH\_DST]};

\item Set the {\em update-scope} to the Ethernet source address, via an \texttt{OFPT\_STATE\_MOD} with command field set to \texttt{OFPSC\_SET\_U\_EXTRACTOR}, field count set to 1, and field list set to \texttt{[OFPXMT\_OFB\_ETH\_SRC]};

\item Assuming a switch with N ports, send  the following $(N+1) \times N$ \texttt{OFPT\_FLOW\_MOD} messages:

\vspace{-9pt}
\begin{footnotesize}
\begin{verbatim}
for i in [PORT1, PORT2, ..., PORTN]:
  flow_mod([match=[state=0, inport = i], 
      action=[FLOOD, set-state(i)]])                            
  for j in [PORT1, PORT2, ..., PORTN]:
    flow_mod([match=[state=j, inport = i], 
        action=[OUTPUT(j), set-state(i)]])
\end{verbatim}
\end{footnotesize}

The MAC learning XFSM table will be configured with the  $(N+1) \times N$ entries - $i,j \in (1,N)$:

\begin{footnotesize}
\begin{tabular}{|l|l|} \hline
{\em match} 			& {\em action} 				\\ \hline
state=DEF, in\_port=i 	& set\_state(i), FLOOD 		\\ \hline
state=j, in\_port=i 	& set\_state(i), OUTPUT(j) 		\\ \hline
\end{tabular}
\end{footnotesize}

\end{enumerate}

The above configuration uses as state labels the port numbers. A lookup to the State Table using the Ethernet destination address will retrieve the port to which the packet should be forwarded, or a DEF (default) state if the address is not stored in the table. The state label (switch output port) and the ingress port are provided as input to the XFSM table, which returns a flood action on all switch ports in the case of default state, or a forward action to the port returned as state otherwise. In both cases, a set\_state(ingress port) instruction will be triggered, to update the entry corresponding to the differing update-scope (the Ethernet source address).

{\bf SW Performance insights}. We have used the MAC learning example to grasp some insights on the performance of our software implementation. We remark that the original OpenFlow implementation we started from (softswitch) is developed in user space and is poorly performing. Therefore, throughput measurements would give us very limited insights, as the processing time is clearly bottlenecked by the packet capture syscall. To better understand if OpenState yields performance impairments with respect to the original OpenFlow implementation, a more appropriate performance measure is the average processing time of {\em each} OpenState primitive. The results reported below refer to a scenario of 50 hosts connected to a software OpenState deployed in a virtualized enviroment (mininet), and configured as a MAC learning switch, with XFSM table and state table containing 2550 and 50 entries, respectively. 

\vspace{3pt}
\begin{footnotesize}
\noindent
\begin{tabular}{ | c | c | c | c | c | }
\hline
\textbf{parsing} & \textbf{s. lookup} & \textbf{xfsm} & \textbf{s. update} & \textbf{actions}\\
\hline
1.14 us& 2.07 us& 804.49 us& 2.23 us& 0.83 us\\
\hline
\end{tabular}
\end{footnotesize}

As expected the processing overhead introduced by the Openstate primitives is negligible (in this case less then 1\%) and dominated by the XFSM table match, i.e. the wildcard match whose performance rapidly becomes a bottleneck for SW implementations as the table size grows (unlike in the TCAM HW case). However, in SW, support for parametric actions is straightforward (whereas in HW they require non trivial techniques such as those recently introduced in \cite{Bos13}). We repeated the measurements by implementing the \texttt{OUTPUT(j)} action as parametric, using as parameter the retrieved state label. In this case, the XFSM table size can be deployed with only $N+1$ entries, and the xfsm transition time descreases from more than 800 us to 13.2 us. 

{\bf Learning on ``unconventional'' flow identifiers}. As the MAC learning example has shown, the addresses used by the learning approach have been configured via the lookup and update scope, and the forwarding to switch ports has been performed accordingly to state labels. By changing the definition of states and the lookup/update scopes, the same learning construction can be cast into widely different scenarios. As an example, consider a data center network where a subset of edge switches, directly connected to the end hosts via {\em edge ports}, act as ingress/egress nodes and are connected each other via a core network engineered via MPLS paths. Core switches are connected to edge swicthed via {\em transport ports}. Once a packet, say a layer 2 Ethernet frame, arrives to an ingress switch, the ingress node must identify which path brings the packet to the egress switch, before adding the appropriate MPLS label and forward the packet on the relevant path. A frequently recurring idea is to use an indentifier of the edgress switch as label itself. The problem, ordinarily solved by dedicated control protocols or by a centralized SDN controller, is to create and maintain a mapping between a destination host address and the relevant egress switch identifier (MPLS label).

At least in principle (because of the obvious emerging inefficiencies), it could be possible to get rid of any control protocol, and adopt the following simple solution mimicking a (hierarchical) layer 2 operation. Each ingress switch maintains a forwarding database mapping destination MAC addresses to egress switch identifiers. When a packet arrives at an ingress switch, a state table is queried using as key its MAC destination; if an entry if found, a label comprising both the egress and the ingress switch identifiers is added to the packet; the core network will forward the packet based on the egress switch identifier. Conversely, if no entry is found in the forwarding database, the packet is broadcast to all the egress switches\footnote{Also the broadcasting will rely on labels, so that core switches need only to maintain forwarding states for egress switch labels rather than per MAC addresses}. In turns, egress switches could use the received packets to learn the mapping between the source MAC address of the packet, and its ingress switch (whose identifier is included the MPLS label). Note that the egress switches will also need to store the ordinary layer 2 forwarding information, i.e. through which edge port the packet shall be forwarded.

Such operation is trivially implemented with OpenState by an XFSM program conceptually very similar to that used for MAC learning, but cast in the new hierarchical port/label setting; specifically, it would suffice\footnote{We haven't actually implemented it, as it is conceptually equivalent to the MAC learning example and would give limited extra insights; indeed we {\em do not claim} that such approach is effective (at least as presented here), but we use it as an hopefully compelling example of how OpenState permits to deploy learning in different settings.} to set (notations in the tables being self-explaining):
\vspace{-6pt}
\begin{enumerate}\setlength{\itemsep}{-2pt} 
\item {\em lookup-scope} = MAC destination;
\item {\em update-scope} = MAC source;
\item state label = the pair $[P_i, S_j]$ with $P_i$ an edge port of the local ingress/egress switch, and $S_i$ being an egress switch identifier;
\item XFSM table - entries for packet incoming from an edge port (outbound packets, the packet is labeled and forwarded to the egress switch(es); the state table learns the edge port of the arriving packet):

\begin{footnotesize}
\begin{tabular}{|l|l|} \hline
{\em match} 					& {\em action} 									\\ \hline
state=[*,DEF], in\_port=$P_i$ 	& set\_state($P_i$,*), Flood 					\\ \hline
state=[*,$S_j$], in\_port=$P_i$ & set\_state($P_i$,*), Fwd $S_j$ 		\\ \hline
\end{tabular}
\end{footnotesize}

\item XFSM table - entries for packet incoming from a transport port (inbound packets, the packet is decapsulated and forwarded to the edge port(s), the state table learns the ingress switch from where the packet arrives):

\begin{footnotesize}
\begin{tabular}{|l|l|} \hline
{\em match} 						& {\em action} 									\\ \hline
state=[DEF,*], label=$S_j$ 	& set\_state(*,$S_j$), Flood 			\\ \hline
state=[$P_i$,*], label=$S_j$ & set\_state(*,$S_j$), Fwd $P_i$ 		\\ \hline
\end{tabular}
\end{footnotesize}

\end{enumerate}

\subsection{DDoS mitigation}

\begin{figure}[t]
\centering
\includegraphics[width=0.47\textwidth]{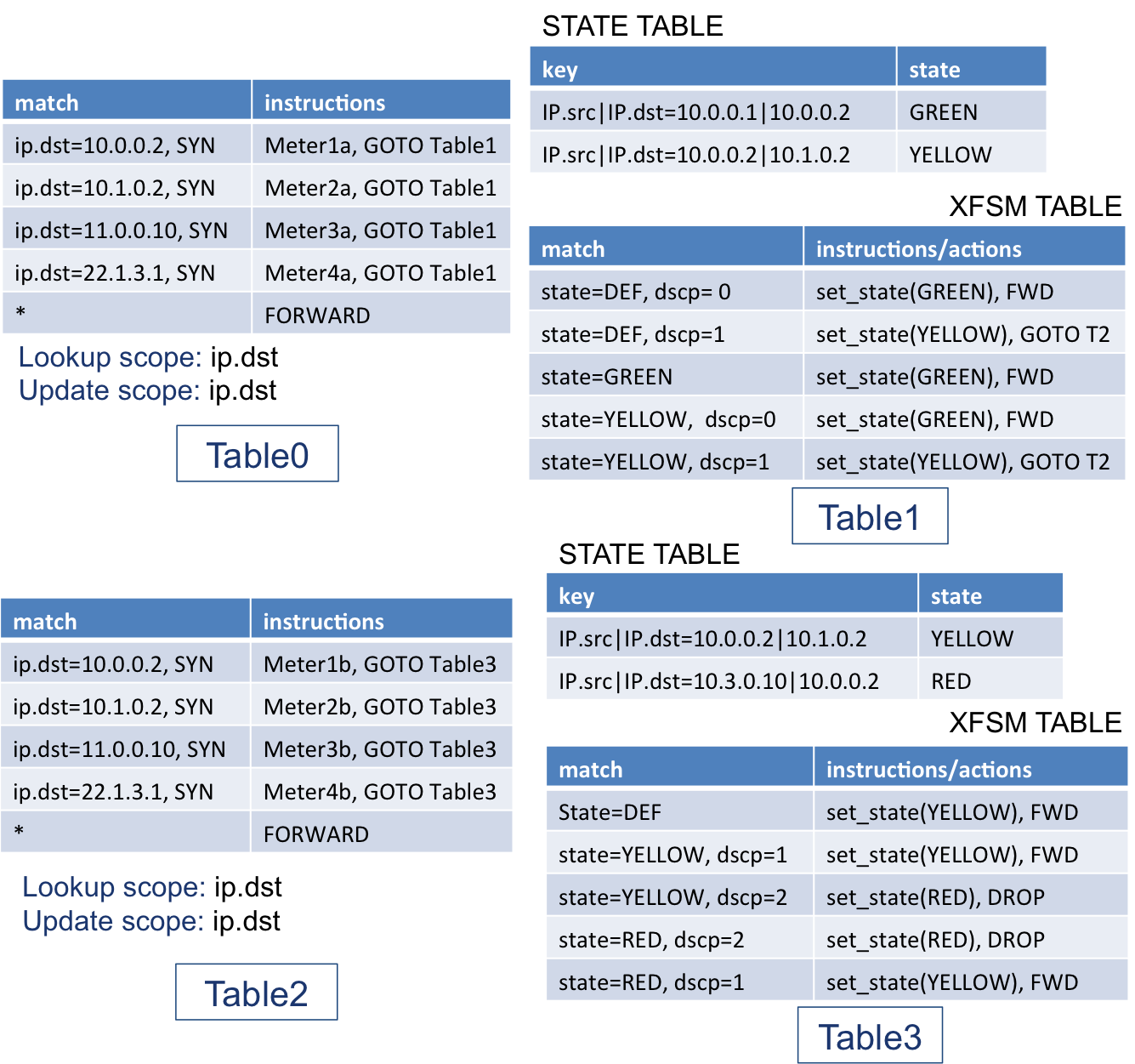}
\caption{DDoS detection use case}
\label{fddos}
\end{figure}

In this section we describe a simple DDoS detection and mitigation mechanism. The proposed use case is not meant to introduce a novel security algorithm but it rather demonstrates a basic OpenState capability that could not be offered by the standard stateless OpenFlow data plane without requiring to forward each new connection to a controller, i.e. the ability of identifying flows generated before and after a given event (an attack in this case). 

This application performs 2 monitoring stages. In Stage1, the switch monitors the bit-rate of incoming TCP SYN packets addressed to a finite list of possible destinations and keep a state for each source/destination IP flows. For each new flow, identified by the pair (IP.src,IP.dst), the switch acts according the following strategy: if a new flow is addressed to a destination for which the SYN bit-rate is under a given threshold, the flow is marked as \texttt{GREEN} and forwarded through the proper switch output port. If instead a new flow is addressed to a destination for which the meter threshold is exceeded, the flow is marked as \texttt{YELLOW} and forwarded to a second monitoring stage. From this point on, all new connections to the same IP address will be forwarded to a second monitoring stage, while all \texttt{GREEN} flows (i.e. those generated before the actual attack) are forwarded. The second monitoring stage (Stage2) is analogous to the first one. If for a given IP address a second SYN-rate threshold is exceeded, all new flows addressed to this IP address are dropped. 

Figure \ref{fddos} describes the actual OpenState implementation of the simple mechanism described above, which is realized with a 4 tables pipeline:  Table0 and Table1 implement stage1; Table2 and Table3 implement Stage2. 

Table0 and Table2 are configured to measure the SYN rate toward a predefined set of IP destinations and are instantiated at startup by the controller 
%
%
using OpenFlow DSCP meters\footnote{It is worth noting that the DSCP field is used to propagate information between tables. A much cleaner solution might be based on a new meter type able to write metadata}. As result of these meter instructions, all flows exceeding the given threshold and burst size will be marked with a DSCP field set to 1 or 2 by respectively Table0 and Table2. 

Table1 implements the actual XFSM for Stage1. 
For each new flow (state \texttt{DEFAULT}) the switch checks for the DSCP field. If DSCP = 0 (meter band under threshold), the flow is marked with state \texttt{GREEN} and forwarded (entry 1). If DSCP = 1 (meter band threshold exceeded), the flow is marked with state \texttt{YELLOW} and the packet is pipelined to Table2 (entry 2). All packets belonging to flows marked as \texttt{GREEN} (entry 3) are forwarded and kept as \texttt{GREEN} (and do pass through Stage2). For each packet belonging to a flow marked as \texttt{YELLOW}, the switch checks for the DSCP field. If DSCP = 0 (meter band rolled back under threshold), the flow state is set to \texttt{GREEN} and the packet is forwarded (entry 4). If DSCP = 1 (meter band still over threshold), the flow is marked with state \texttt{YELLOW} and the packet is pipelined to Table2 (entry 5).

Table3 implements the actual XFSM for Stage2. 
If the table receives a packet with DEFAULT state (i.e. the first packet of a flow pipelined to table3), the flow is marked as \texttt{YELLOW} and the packet is forwarded, regardless the DSCP field value (entry 1). For each packet belonging to a flow marked as \texttt{YELLOW}, the switch checks for the DSCP field. If DSCP = 1 (second meter band under threshold), the flow state is kept to \texttt{YELLOW} and the packet is forwarded (entry 2). If DSCP = 2 (second meter band threshold exceeded), the flow state is set to \texttt{RED} and the packet is dropped (entry 3). Finally, for all packets marked as \texttt{RED} the switch checks for the DSCP field. If DSCP = 2 (meter band still over threshold) the packet is dropped and the flow state is kept unchanged (entry 4). If DSCP = 1 (meter band back under threshold) the packet is forwarded and the flow state is rolled back to \texttt{YELLOW} (entry 5).

\subsection{Forwarding consistency}
\label{s:forwarding-uc}

Load balancing is another interesting use case where states are necessary to ensure consistency in forwarding decisions for packets of a same transport layer flow. In current OpenFlow this is handled via group tables and ''switch-computed selection algorithm'' whose configuration and state management are external to OpenFlow. OpenState can be used to configure and execute the logic of forwarding consistency in the switches. We show here three examples where simple algorithms can be implemented. 

{\bf One-to-Many load balancing}
 This is the case of a switch forwarding packets from one input port to multiple output ports. For example, the first packet of a TCP connection from host $h_a$ to $h_b$ arrives at the input port $p_1$, the switch has to choose on which of the many output ports the packet must be forwarded, for example $p_j$. We want that all the next packets of the same TCP connection to be forwarded on the same $p_j$ port.

\begin{figure}[t]
\centering
\includegraphics[width=0.40\textwidth]{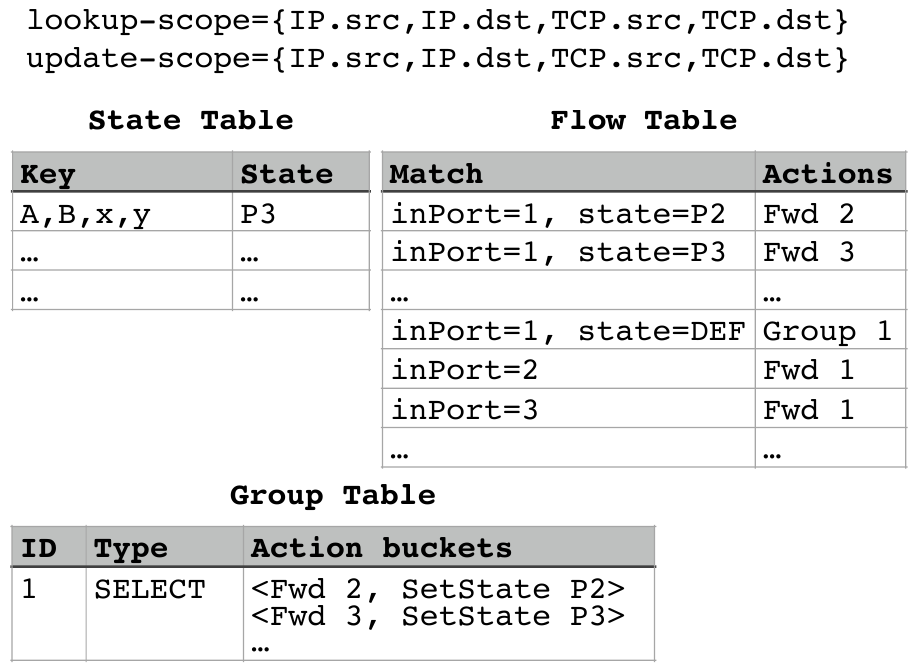}
\caption{One-to-Many load balancer implemented using our OpenState extension of OpenFlow}
\label{fig:1tomany}
\end{figure}

Figure~\ref{fig:1tomany} shows a simplified implementation using OpenState. Here the lookup-scope associated with the state table is used to distinguish between different TCP connections. When the first packet enters the switch there's no previous known state associated, a \texttt{DEFAULT} state is then appended to the packet header. All packets entering the only input port and matching the \texttt{DEFAULT} state are processed by a group table. A group entry of type \texttt{SELECT} is used to perform the actual load balancing. From the OpenFlow specification v1.3+ we know that when a group entry is of type \texttt{SELECT}, only one bucket in the group will be selected by a a switch-computed algorithm and executed. What we provide with OpenState is a learning process for the selected port. Indeed, for each bucket execution, along with forwarding the packet we store the output port by calling a SetState. The next packet of the same connection will then match a not \texttt{DEFAULT} state and will be forwarded accordingly. Note that state lookup timeout with rollback to DEFAULT or explicit connection termination events can be used to flush unused entries.

{\bf Many-to-One Reverse Path Forwarding}
This case is complementary to previous one with forwarding consistency that must be ensured according to packets received in the reverse direction (many-to-one). For example, the first packet of a TCP connection from host $h_a$ to $h_b$ arrives at one of the many input ports, for example $p_i$. The switch forwards the packet on the only output port $p_1$. We want all packets of the {\em reverse flow}, namely all packets from $h_b$ to $h_a$ of the same TCP connection, to be forwarded on the same ingress port $p_i$.


With respect to previous case, here lookup-scope and update-scope are different and have reversed IP/TCP src and dst fields. 
When the first packet of a TCP connection (from IP $A$ to $B$, from TCP port $x$ to $y$) enters the switch from input port $p_i$. No row is matched in the state table, a DEFAULT state is appended. The packet is forwarded to the only output port $p_1$ while the ingress port is stored by calling a SetState. 
The state table is updated with the corresponding key. From now on all the reverse packets (from IP $B$ to $A$, from TCP port $y$ to $x$) will be forwarded to port $p_i$. As in previous case, state timeouts can be used to flush entries for closed connections.

{\bf Many-to-Many}
Combining the first two we obtained another particularly significant case.  We want here to load balance on the output ports while doing reverse path forwarding on the input ports. This could be for instance the case of a switch forwarding packets between aggregated links. For example, the first packet of a TCP connection from host $h_a$ to $h_b$ arrives at one of the many input ports $p_i$. The packet is forwarded to one of the many output ports $p_j$ (as in the One-to-Many case a SELECT group entry is used to select a random port). From now on all the packets of the same TCP connection from $h_a$ to $h_b$ must be forwarded to $p_j$, while all the packets of the reverse flow from $h_b$ to $h_a$ must be forwarded to $p_i$. To achieve this result we need to update 2 entries of the state table: (i) the output port $p_j$ and (ii) input port $p_i$ for the packets of the reverse flow from $h_b$ to $h_a$. This could be achieved by performing 2 SetState using 2 different update-scopes. This possibility is actually not implementable using the current OpenState primitives because we actually allow updates on only one update-scope. However, the extension to (a small number of) multiple state updates targeting different flows is rather straightforward and does not affect the basic assumptions of the OpenState abstraction.

\section{Related work}

OpenFlow, now at version 1.4 \cite{OF1.4}, has undergone several extensions (more flexible header matching, action bundles, pipelined tables, synchronized tables, multiple controllers, and many more), and new important ones are currently under discussion, including typed tables \cite{FAWG} and flow states \cite{Cra13}. The need to rethink OpenFlow data plane abstraction has been recently recognized by the research community \cite{Bos13, Bos13b, Son13}. In \cite{Bos13}, the authors point out that the rigid table structure of current hardware switches limits the flexibility of OpenFlow packet processing to matching on a fixed set of fields and to a small set of actions, and propose a logical table structure, RMT (Reconfigurable Match Table), on top of the existing fixed physical tables and new action primitives. Notably, the proposed scheme allows not only to consider arbitrary width and depth of the matching for the header vector but also to define actions that can take input arguments and rewrite header fields. Along the same line but with a more general approach, in \cite{Bos13b} the authors propose P4, a high-level language to program packet processors which focus on protocol-independence. P4 allows the programmer to define packet parser (able to support matching on new header fields) and to implement arbitrary parametric actions as the composition of reusable primitives for packet header manipulation. It must be said that P4 is presented as a straw-man proposal for how OpenFlow should evolve in the future, rather than a proof-of-concept. In \cite{Son13} a  Protocol-Oblivious Forwarding (POF) abstraction model is proposed as a set of low-level Forwarding Instruction Set (FIS) (comparable to those found in the assembly language). Especially, the authors propose new stateful instructions to actively manipulate the flow table entries. Even though prototype implementations (hardware and software-based) are presented, POF can be considered a clean slate proposal with substantial differences from existing SDN programming abstractions and implementations that can hardly be put into an evolutionary perspective.

In \cite{Jey13}, the approach is even more radical and, similarly to the early work on active networks, packets are allowed to carry a tiny code that define processing in the switch data plane. A very interesting aspect is the proposal of targeted ASIC implementations where an extremely small set of instructions and memory space can be used to define packet processing. 

The use of XFSMs advanced features was initially inspired by \cite{Tin12} where (bytecoded) XFSMs were used to convey a desired medium access control operation into a specialized (but closed \cite{Bia12}) wireless interface card. While the abstraction (XFSM) is similar, the context (wireless protocols versus flow processing), technical choices (state machine execution engine versus table-based structures), and handled events (signals and timers versus header matching), are not nearly comparable.

We can also find FSM models at the basis of the approach to define network policies in PyResonance \cite{pyresonance}. However, FSMs in PyResonance are defined at the controller level and later translated to OpenFlow rules using a reactive approach.

Finally, while completing the writing of this paper, we noticed \cite{moshref14b} in the announced program of HotSDN 2014, and got in contact with the authors who kindly shared their submitted version. The FAST approach therein proposed shares with OpenState the basic idea of using state machines for modifying the forwarding behavior of switches (although it does not support our proposed cross-flow state handling), and describes a software prototype based on Open vSwitch. The fact that our basic idea is being independently emerging elsewhere supports our belief that the introduction of stateful approaches in the OpenFlow data plane might be crucial. Concerning technical aspects, FAST defines a high level programming abstraction that makes use of variables and functions to define events and transitions, whose HW implementation may be not trivial. Indeed, as also anticipated in our earlier version of this paper \cite{ccr14}, we believe that the most compelling way to show that a proposed approach is compatible with OpenFlow hardware and can operate at wire speed is an actual hardware implementation, which we provided in this paper as proof-of-concept, and which permitted us to gain insights on possible (but luckily minor) pipelining limitations emerging from the backward loop that both our OpenState as well as FAST do introduce.


\section{Conclusions}

OpenState is a first attempt to permit platform-agnostic programming of a subset of stateful control functions directly inside the switches, while retaining wire-speed performance. In adherence with the OpenFlow strategy, we take a pragmatic approach as well, by restricting control functions to the ability to dynamically install and evolve flow states, and associate different forwarding actions to different combinations of flow states and packet header matching events. Our main, and perhaps surprising, finding is that any control function that can be modelled as a Mealy Finite State Machine (provided it uses the same set of actions supported in OpenFlow, of course), is readily deployable using commodity OpenFlow hardware and can be developed as a very simple extension of OpenFlow version 1.3. With very simple hardware extensions, OpenState can further support ``cross-flow'' state handling, i.e. permit the arrival of a packet of a given flow to trigger a state transition for a different flow. The viability of OpenState was demonstrated by two (HW and SW) proof-of-concept implementations, and example OpenState applications were provided. 

The road towards wire-speed platform-agnostic ``full'' control plane programmability is clearly ambitious and long, and OpenState is just a start which, we hope, may stimulate interest and discussion in our community. We foresee at least three research questions which require further insights: i) which {\em architectural} evolutions of OpenState can support more flexible programming abstractions, such as a complete extended finite state machine \cite{Che93}? ii) how to extend the set of OpenFlow actions and switch-level states to support furter control tasks? And how the ability to offload control programs directly inside the switches may influence broader SDN frameworks?

\newpage

\bibliographystyle{IEEEtran}

\bibliography{biblio}

\end{document}